\documentclass[prd,aps]{revtex4}
\usepackage{graphicx}
\usepackage{amsmath}
\usepackage{amsfonts}
\usepackage{amssymb}
\DeclareGraphicsExtensions{.pdf,.png,.jpg}

\begin{document}
\preprint{{hep-th/yymmnnn} \hfill {UCVFC-DF-17-2005}}
\title{Hamiltonian BF theory and projected Borromean Rings}
\author{ Ernesto Contreras$^{1}$, Adalberto D\'{\i}az$^{1}$ and Lorenzo Leal$^{1,2}$}

\affiliation {1. Centro de F\'{\i}sica Te\'{o}rica y
Computacional, Facultad de Ciencias, Universidad Central de
Venezuela, Apartado Postal 47270, Caracas 1041-A, Venezuela. \\
 2. Departamento de F\'{\i}sica, Universidad Sim\'on Bol\'{\i}var, Apartado Postal 89000,
Caracas 1080-A, Venezuela.\\ }

\begin{abstract}
 It is shown that the canonical formulation of
the abelian BF theory in $D=3$ allows to obtain topological invariants associated to curves and points in the plane. The method consists on finding the Hamiltonian on-shell of the theory coupled to external sources with support on curves and points in the spatial plane. We explicitly calculate a non-trivial invariant that could be seen as a "projection" of the Milnor's link invariant $\overline{\mu}(1,2,3)$, and as such, it measures the entanglement of generalized (or projected) Borromeans Rings in the Euclidean plane.

\end{abstract}

\maketitle


The obtention of knot and link invariants through the study of topological gauge theories can be traced back to the calculation of the vacuum expectation value  of the Wilson Loop in the
Chern-Simons (CS) theory \cite{witten}. When the gauge fields are integrated out, the resulting expression only depends on the Wilson lines, and, since the theory is metric independent, only the entanglement properties of these lines should matter. That this is what indeed happens  can be shown explicitly in the Abelian case, where the invariant obtained is
the Self-Linking Number or the Gauss Linking Number (GLN),
depending on whether one deals with one or several Wilson lines. Also, in the non-Abelian case, analytical expressions for knot or link invariants can be obtained by perturbative methods \cite{guadagnini}, and using general arguments it can be shown that the exact Wilson loop average is an invariant  related to knot or link polynomials, such as the  Jones polynomial  \cite{witten,guadagnini,labastida}.

Beside this quantum method, there is also a classical approach to obtain knot and link invariants from topological gauge theories, that has been described elsewhere \cite{lorenzo2,lorenzo1, lorenzo3,rafael}. The method consists in solving
the classical equations of motion in order to calculate the  action on-shell of
the topological theory coupled to external currents with support on closed curves. The
action  on-shell so obtained is a functional that only depends on the curves, and, since the theory is
metric independent, it yields analytical formulae for knot or link invariants.  This method can be rigorously stated \cite{rafael}, and can also be extended to deal with models
where the symmetry group is other than the group of
diffeomorphisms \cite{rafael, area}.
Using this method one is able to calculate, for instance, an analytical expression for the Milnor's Linking Coefficient
$\overline{\mu}(1,2,3)$ , which is an invariant that follows the Gauss Linking Number
in an infinite sequence of link invariants discovered by Milnor some years ago \cite{milnor}. This invariant serves to detect the braiding of the Borromean Rings, which is a well known link that comprises
three curves whose Gauss Linking Numbers vanish, although they are certainly entangled
\cite{milnor,rolfsen,mellor}.

 In this paper we address the following question: which is the canonical or Hamiltonian counterpart of the method mentioned above? Stated in other words, is there a kind of "remnant" of link invariants when one "reduces" the topological theory in $D$ space-time dimensions  to the $D-1$-dimensional space by passing from the Lagrangean to the Hamiltonian formulation in the classical  theory? Since the scope of this paper is to set up a preliminary study, we face this question by considering a simple topological field theory: the Abelian BF model coupled to external sources. As we shall see, even in this simple model already appear issues of what we are looking for. In fact, we find that with the simplest choice for the sources that one could conceive, the Hamiltonian on-shell yields the winding number of a closed curve (a static "electric current") around a point (a static "charge"), which is certainly a topological invariant. Moreover, for a more involved choice for the sources (see equations (18-21)), we found an invariant that measures the entanglement properties of a closed curve with respect to two points, and that has the interesting property of vanishing when one of the points is dropped. This fact mimics what occurs, in three dimensions, with the Borromean Rings, which can be disentangled if one drops just one of them. In this sense, one could say that the expression obtained in the case we are going to study can be seen as a "projection" of the Milnor Invariant mentioned above.

 The action of the BF theory with external sources  $J^{\mu}$ , $K^{\mu}$, is given by

 \begin{equation}
 S = \int d^{3}x (\,\varepsilon^{\mu\nu\rho}B_{\mu}\partial_{\nu}A_{\rho}  -J^{\mu}B_{\mu} -K^{\mu}A_{\mu}).  \label{1}
\end{equation}
The theory is gauge-invariant whenever the currents are conserved; also, it is invariant under space-time diffeomorphisms, provided that the currents transform as vectorial densities under general coordinate transformations. Moreover, since the metric does not appear in the action, the theory is topological. This fact is the basis for obtaining topological invariants both within the classical and quantum frameworks.

The canonical formulation of this theory  can be summarized as follows. After space-time decomposition

\begin{equation}
 S = \int d^{3}x (\,\varepsilon^{ij} (\partial_{t}A_{i}) B_{j}  -J^{i}B_{i} -K^{i}A_{i} + B_{0}(\varepsilon^{ij} \partial_{i}A_{j}-J_{0})
 + A_{0}(\varepsilon^{ij} \partial_{i}B_{j}-K_{0}))  ,\label{2}
\end{equation}
and noticing that the Lagrangean is  first order in time derivatives,
we can read the fundamental equal-time Poisson brackets

\begin{equation}
  \{ A_{i}(x) , \varepsilon^{jk}B_{k}(y) \}= \delta _{i} ^{j} \delta ^{2} (x-y) . \label{3}
\end{equation}
The fields $A_{0}$ $B_{0}$ are not properly dynamical fields, but Lagrange multipliers that enforce the (spatial) gauge constraints which are given by

\begin{equation}
 \varepsilon^{ij} \partial_{i}A_{j}-J_{0} = 0 ,\label{4}
\end{equation}
and
\begin{equation}
 \varepsilon^{ij} \partial_{i}B_{j}-K_{0} = 0 . \label{5}
\end{equation}
The remaining equations of motion (i.e., those which involve time derivatives) can be easily obtained through the Poisson brackets of the canonical variables with the Hamiltonian, as usual. However, in order to compute the Hamiltonian on-shell it is not necessary to integrate those equations, since the "physical" (or first class, in Dirac sense) Hamiltonian is just given by

\begin{equation}
 H = \int d^{2}x (J^{i}B_{i} + K^{i}A_{i}) \label{6}.
\end{equation}
 Then, by integrating the constraint equations (\ref{4})  and (\ref{5}), and substituting the spatial components of the currents into (\ref{6}) the Hamiltonian on-shell can  be readily computed. Since the Hamiltonian retains the topological character of the action, one should expect that the Hamiltonian on-shell also yields formulae for "projected knot invariants" in the $2$-dimensional space, inasmuch the action on-shell does with ordinary knot invariants.  Let us see how this works in two concrete examples. We shall take, as our first case, the following set of conserved currents

\begin{equation}
  J^{\mu} (x)= (J_{0}, J^{i}) = (0, \oint _{\gamma} dy^{i} \delta ^{2} (x-y))  ,\label{7}
\end{equation}

\begin{equation}
  K^{\mu}(x)= (K_{0}, K^{i}) = (\delta ^{2} (x-x_{0}), \overrightarrow {0}))  ,\label{8}
\end{equation}
where $x_{0}$ and $\gamma$ are a fixed point and a closed curve in the plane respectively. If desired, these currents could be thought as corresponding to a "static point charge" and a "static current" associated to the gauge fields, but since we are not dealing with the Maxwell field this correspondence should not be taken literally.
Integrating out the constraint (\ref{5}) we find

\begin{equation}
B_{i}(x)= \frac{1}{2\pi}\varepsilon_{ij} \frac{(x-x_{0})^{j}}{\|x-x_{0} \|^{2}} + \partial_{i}\Lambda  , \label{9}
\end{equation}
where $\Lambda$ is an arbitrary function of the spatial coordinates.
Plugging this result in (\ref{6}) we finally obtain the Hamiltonian on-shell

\begin{equation}
 H_{OS}= \frac{1}{2\pi} \oint _{\gamma} dx^{i} \varepsilon_{ij} \frac{(x-x_{0})^{j}}{\|x-x_{0} \|^{2}} . \label{10}
\end{equation}
This formula measures the winding number of the curve $\gamma$  around  the point $x_{0}$. This is the simplest topological invariant associated to a closed curve and a point in the plane. It should be seen as the planar counterpart of the (three dimensional) Gauss linking number between two closed curves: the point may be thought as the intersection of one  curve  living in the "missing" third dimension, with the plane where the other curve lies.

Inspired by this simple result, let us now try a more interesting choice for the currents. First, we find it convenient to introduce the following definitions. Associated with a region $\Sigma$ in the plane, let us define the $0$-form $f^{(\Sigma)}(x)$ (or  just $f^{(\Sigma)}_{x}$) as

\begin{equation}
f^{(\Sigma)}(x)=\int\limits_{\Sigma}d^{2}x'\delta^{2}(x-x'). \label{11}
\end{equation}
Also, we define the $1$-form  $g_{i}^{(\gamma)}(x)$ (sometimes we shall use $g_{ix}^{(\gamma)}$) associated to a curve $\gamma$  as

\begin{equation}
g_{i}^{(\gamma)}(x) = \varepsilon_{ij}T^{ (\gamma)j}(x)=\varepsilon_{ij} \int_{\gamma} dx'^{j} \delta^{2} (x-x'). \label{12}
\end{equation}
In passing, we have introduced  the "path-coordinate" or
"form-factor" $T^{(\gamma)j}(x)$ associated to the path $\gamma$
in this expression (we shall also use the notation
$T^{(\gamma)jx}$). This object is the first member of an infinite
family of "path-coordinates"  that serve to characterize
non-parametric curves in a general manifold \cite{dibartolo}. The
second member is the "two-points path-coordinate"
\begin{equation}
T ^{(\gamma)ij}(x,y)= T ^{(\gamma)ix,jy} =\int_{\gamma}dz^{i}\int\limits_{x_{0}}^{z}dz'^{j}\delta^{2}(x-z)\delta^{2}(y-z') , \label{13}
\end{equation}
that obey the differential constraints

\begin{eqnarray}
\partial_{ix} T^{(\gamma)i x,j y}&=&(-\delta^{2}(x-x_{0})+\delta^{2}(x-y))T^{(\gamma)j y},\\ \label{14}
\partial_{jy}T^{(\gamma)i x,j y}&=&(\delta^{2}(y-x_{0})-\delta^{2}(y-x))T^{(\gamma)i x}.    \label{15}
\end{eqnarray}
Here, $x_{0}$ is the beginning point of the curve. The former expressions hold when the curve is closed.  Similar constraints holding for the general case can be given if needed. There is also an algebraic constraint
\begin{equation}
 T^{(\gamma)(i x,j y)}=\frac{1}{2}(T^{(\gamma)i x,j y}+T^{(\gamma)j y,i x})=T^{(\gamma)i x}T^{(\gamma)j y}\ ,\label{16}
\end{equation}
which shows that only the antisymmetric part of $T^{(\gamma)i x,j y}$ is independent of the form-factor. It can be easily shown that the forms defined above also obey

\begin{eqnarray}
-\partial_{i}f^{(\Sigma)}(x)=\oint\limits_{\partial\Sigma}\varepsilon_{ij} \, dx'^{j}\delta^{2}(x-x')=g^{(\partial\Sigma)}_{i}(x) \label{17}
\end{eqnarray}
where the curve $\partial\Sigma$ is the boundary of  $\Sigma$ .

With these tools at hand  let us go to our second example. We take two open curves  $\gamma_{1}$ and $\gamma_{2}$ beginning at the spatial infinity and ending at points  $x^{(1)}_{f}$ and $x^{(2)}_{f}$ respectively, and a closed curve $\gamma_{3}$ surrounding a region $\Sigma_{3}$. Using these ingredients we form the currents

\begin{eqnarray}
K^{0}(x)&=&\delta^{2}(x-x^{(1)}_{f})-\delta^{2}(x-x^{(1)}_{0})=-\partial_{i}T^{(\gamma_{1})i x} \\\label{18}
K^{i}(x)&=&0\\\label{19}
J^{0}(x)&=&0\\\label{20}
J^{i}(x)&=&\varepsilon^{ij}g^{(\gamma_{2})}_{jx}f^{(\Sigma_{3})}_{x}+\int d^{2}yT^{(\gamma_{3})[i x,j y]}g^{(\gamma_{2})}_{j y}\label{21}
\end{eqnarray}
that will  replace the old currents coupled to the $A$ and $B$
fields in the $BF$ action (since $x^{(1)}_{i}$, the starting point
of  $\gamma_{1}$, is at the spatial infinity, the second Dirac
delta function in the first line of equation (14) indeed
vanishes). The sources defined in equations (18-21) are
topological, in the sense that do not depend on the metric and
transform covariantly  under general coordinate changes. This can
be easily seen by noticing that the path-coordinates defined in
equations (\ref{12}) and (\ref{13}) transform as local and
bi-local tensorial densities under coordinate transformations
respectively \cite{dibartolo}. This fact is essential to
maintain the topological character of the action and of the
corresponding Hamiltonian.

The canonical equations of motion are formally the same as before.  The conservation of the current $K^{\mu}(x)$ is
 trivial. This is not the case for the other one, as we show. To calculate $\partial_{i}J^{i}=0$ we first
 compute

\begin{eqnarray}
\partial_{i}(\varepsilon^{ij}g^{(2)}_{j}f^{(3)})&=&\partial_{i}(\varepsilon^{ij}g^{(2)}_{j})f^{(3)}+
\varepsilon^{ij}g^{(2)}_{j}\partial_{i}(f^{(3)})\\ \nonumber
&=&\delta^{2}(x-x^{(2)}_{f})f^{(3)}+\varepsilon^{ij}g^{(2)}_{j}\partial_{i}f^{(3)}, \ \label{22}
\end{eqnarray}
that corresponds to the derivative of the first term of $J^{i}$.
In turn, the contribution of the second term is

\begin{eqnarray}
\partial_{i}\int d^{2}yT^{(3)[i x,j y]}g^{(2)}_{j y}&=&-\int d^{2}y(-\delta^{2}(x-x_{0})+
\delta^{2}(x-y))T^{(3)jy}g_{jy}^{(2)}\\ \nonumber
&=&\delta^{2}(x-x_{0}) \, \varepsilon_{jk} \oint_{(3)} dy^{j}\int_{(2)} dw^{k}\delta^{2}(y-w)+\varepsilon_{jk}\partial_{k}f_{x}^{(3)}g_{jx}^{(2)}.\label{23}
\end{eqnarray}
In these calculations we made intensive use of the properties of the forms associated to paths, loops and regions defined before, as well as the properties of the loop coordinates already discussed. Putting all together we finally arrive at the condition
\begin{eqnarray}
\delta^{2}(x-x^{(2)}_{f})f^{(3)}_{x}+
\delta^{2}(x-x_{0}) \, \varepsilon_{jk} \oint_{(3)} dy^{j}\int_{(2)} dw^{k}\delta^{2}(y-w)=0. \label{24}
\end{eqnarray}
This result establishes that the current $J^{\mu}(x)$ is conserved
whenever the curve $\gamma_{3}$ does not enclose the ending point
of the open curve $\gamma_{2}$. In fact, the first term of the
last expression vanishes in that case, since the Dirac function
can not pick any point $x$  lying  inside $\gamma_{3}$ and, at the
same time, coinciding with the end of $\gamma_{2}$. On the other
hand, the second term measures the oriented number of cuts between
$\gamma_{2}$ and $\gamma_{3}$ (times certain Dirac function that
does not matter for the discussion that follows), and this number
also vanishes provided that the curve $\gamma_{2}$ does not end
inside $\gamma_{3}$. As we shall soon see, this result regarding
current conservation has an interesting relationship with a
similar issue about the consistence of the Milnor invariant
mentioned earlier.

As before our goal is to compute the Hamiltonian on-shell. This
will be accomplished  by integrating out the constraints (just
only one of them, as in the former case) and substituting the
currents and the fields (in terms of the currents) into the
Hamiltonian. This time, we find it convenient, instead of
integrating the constraint equation (\ref{5}) as  before (see equation (\ref{9})
), to write the result as
\begin{eqnarray}
 B_{i}(x)&=&\varepsilon_{ji}T^{(\gamma_{1})jx}+\partial_{i}\Lambda \label{25}
\end{eqnarray}
\\
This way of integrating the constraint (\ref{5}) is equivalent to the former, due to the gauge freedom encoded in the gradient that can be added. Whether one chose one or another formula is a matter of convenience. Substituting this expression into the Hamiltonian (\ref{6}) and after some algebra that includes an integration by parts and the use once more of the conservation of the current given in (21), we can finally write down the Hamiltonian on-shell as

\begin{eqnarray}
I=\int d^{2}x\varepsilon^{ij}g^{(1)}_{i}g^{(2)}_{j}f^{(3)}+\int
d^{2}x\int d^{2}y T^{[i x,j y]}_{(3)}g^{(1)}_{i x}g^{(2)}_{j y}.\label{26}
\end{eqnarray}

To discuss the "topological" interpretation of this formula and its relationship with the Milnor
coefficient $\overline{\mu}(1,2,3)$, it will be useful to make the following considerations. We have already
 shown that in order to the equations of motion be consistent, the ending point of $\gamma_{2}$ should be outside
  $\gamma_{3}$. Furthermore, we shall show that the Hamiltonian on-shell is independent of the shapes of the curves
  $\gamma_{1}$, $\gamma_{2}$, and only depends on their ending points, provided that both ending points lie
   outside  $\gamma_{3}$. Since $I$ is antisymmetric in $\gamma_{1}$, $\gamma_{2}$,
    it suffices to prove this for one of the curves, let us say $\gamma_{1}$.  Consider the variation of $I$
    when the curve $\gamma_{1}$ is replaced by another one, say $\gamma_{1}^{,}$, that ends at the same point
     that the former (recall that our open curves always begin at spatial infinity). The composed curve
      $\gamma_{1}^{,} . (-\gamma_{1})$ surrounds a surface, and we have

\begin{eqnarray}
 \Delta g^{(1)}_{i}=g^{(1')}_{i}-g^{(1)}_{i}=\varepsilon_{ij}(T^{(1')j x}-T^{(1)j x}).\label{27}
\end{eqnarray}
But $T^{(1')j x}-T^{(1)j x}$ is the form factor of the composed curve mentioned above, hence
\begin{eqnarray}
\Delta
g^{(1)}_{i}=g^{(\partial\Sigma^{(1)})}_{i}=-\partial_{i}f^{(1)}.\label{28}
\end{eqnarray}
Therefore,  the variation $\Delta I$ is given by

\begin{eqnarray}
\Delta I&=&\int d^{2}x\varepsilon^{ij}\Delta g^{(1)}_{i}g^{(2)}_{j}f^{(3)}+\int d^{2}x\int d^{2}y
T^{(3)[i x,j y]}\Delta g^{(1)}_{i x}g^{(2)}_{j y}\\ \nonumber
&=&\int
d^{2}x(\varepsilon^{ij}f^{(1)}(x)\partial_{i}g^{(2)}_{j}f^{(3)}+\varepsilon^{ij}(f^{(1)}(x))
g^{(2)}_{j}\partial_{i}f^{(3)})\\ \nonumber
&+&\int d^{2}x\int d^{2}y (-\delta^{2}(x-x_{0})+\delta^{2}(x-y))T^{(3)j y}f^{(1)}(x)g^{(2)}_{j y}\\ \nonumber
&=&-f^{(1)}(x^{(2)}_{f})f^{(3)}(x^{(2)}_{f})+ f^{(1)}(x_{0})f^{(3)}(x^{(2)}_{f}),\ \label{29}
\end{eqnarray}
 and vanishes provided that $x^{(2)}_{f}$ is outside $\Sigma_{3}$.
It is easy to see that $I$ is also unchanged  when we
change   $\gamma_{2}$ (instead of $\gamma_{1}$), whenever $x^{(1)}_{f}$ lies outside $\Sigma_{3}$.

Then, the Hamiltonian on-shell $I$ just depends on the ending
points of the curves  $\gamma_{1}$, $\gamma_{2}$, and
the closed curve  $\gamma_{3}$. Now the following question rises:
does $I$ represents a kind of topological invariant associated to  these objects? The
answer is in the affirmative, as can be seen by noticing that the
formula for $I$ is both: a) invariant under diffeomorphisms in the plane,
and b) metric independent. Then it is a
"topological" object, in the same sense that the action on-shell (or the Wilson Loop average in the quantum formulation) of
topological theories is a topological quantity. To get insight
into what  $I$ means, let us calculate it for the geometric
currents of figure (1). Here, the closed curve $\gamma_{3}$
goes around one of the points and then around the other one; after
that, $\gamma_{3}$ "undoes" the path around the first point, and
finally  "undoes" the path around the second one too. It should be
noticed that if one of the points were dropped, $\gamma_{3}$ could
be "unknoted" from the other point, since this curve does not
enclose any of the two points separately. But the order in which
$\gamma_{3}$ goes around the points matters, and it is easy to see
that the hole structure is entangled! For the lector familiarized
with the Borromean rings, it will be clear that the
picture of figure (1) is nothing but a "plane" version of them: the big dots in the picture should be seen as the intersections of curves that cut the plane were
$\gamma_{3}$ lies. The on-shell Hamiltonian $I$ also
represents the "planar version" of the Milnor invariant
$\overline{\mu}(1,2,3)$. In fact, the "two points form factor"
that appears in the current $J^{i}(x)$  just takes into account
that the "winding and unwinding" of $\gamma_{3}$ around the points
is done in such an order that it is not possible to disentangle
the collection of points and curve. Moreover, a careful
integration of $I$ for this picture yields $1$ (or $-1$,
depending on the orientation chosen for plane), while it would vanish, for instance, for the "unknot": a circle and two points outside it.

\begin{figure}
\begin{center}
\includegraphics[width=7cm,height=4cm]{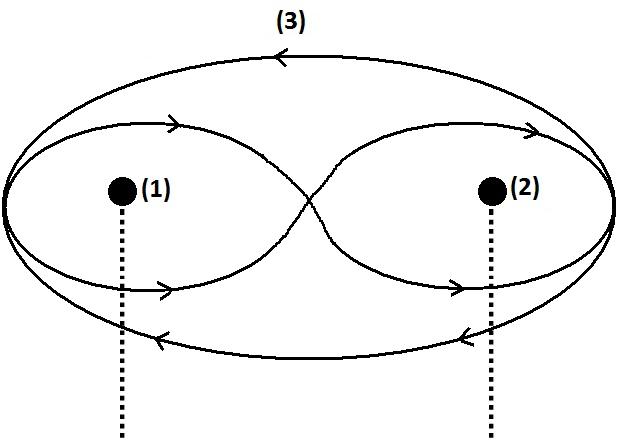}
\caption{Planar "Borromean Rings"}
\label{88}
\end{center}
\end{figure}

As we saw, the invariant $I$ is well defined when the ending points of the auxiliary curves $\gamma_{1}$, $\gamma_{2}$ do not lie inside the region surrounded by $\gamma_{3}$. Otherwise, the equations of motion from which the Hamiltonian on-shell derives become inconsistent  (one of the currents fails to be conserved). It should be remembered that this feature is also present in the three-dimensional case: the Milnor coefficient $\overline{\mu}(1,2,3)$ is well defined when the Gauss Linking Numbers of each pair of curves vanish \cite{milnor}. This completes the analogy between both invariants, and for all these reasons $I$ should be considered a two-dimensional version of the Milnor coefficient $\overline{\mu}(1,2,3)$.




\end{document}